\documentclass[aps,pra,showpacs,superscriptaddress,preprint]{revtex4}
\usepackage{amsfonts}
\usepackage{amssymb}
\usepackage{amsmath}
\usepackage{graphicx}

\catcode`ð=\active
 \defð{\u{g}}
 \catcode`Ð=\active
 \defÐ{\u{G}}
 \catcode`Ý=\active
\defÝ{\. I}
 \catcode`ö=\active
\defö{\"{o}}
 \catcode`Ö=\active
 \defÖ{\"O}
 \catcode`ü=\active
 \defü{\"{u}}
 \catcode`Ü=\active
 \defÜ{\"{U}}
 \catcode`Þ=\active
\defÞ{\c{S}}
 \catcode`þ=\active
 \defþ{\c{s}}
 \catcode`ý=\active
 \defý{{\i}}
 \catcode`ç=\active
\defç{\d{c}}
 \catcode`Ç=\active
\defÇ{\d{C}}

\begin{document}

\title{Approximate bound states of the Dirac equation with some physical
quantum potentials}
\author{Sameer M. Ikhdair}
\email[E-mail: ]{sikhdair@neu.edu.tr}
\affiliation{Physics Department, Near East University, Nicosia, North Cyprus, Turkey}
\author{Ramazan Sever}
\email[E-mail: ]{sever@metu.edu.tr}
\affiliation{Physics Department, Middle East Technical University, 06531, Ankara,Turkey}
\date{%
\today%
}

\begin{abstract}
The approximate analytical solutions of the Dirac equations with the
reflectionless-type and Rosen-Morse potentials including the spin-orbit
centrifugal (pseudo-centrifugal) term are obtained. Under the conditions of
spin and pseudospin (pspin) symmetry concept, we obtain the bound state
energy spectra and the corresponding two-component upper- and lower-spinors
of the two Dirac particles by means of the Nikiforov-Uvarov (NU) method in
closed form. The special cases of the $s$-wave $\kappa =\pm 1$ ($l=%
\widetilde{l}=0)$ Dirac equation and the non-relativistic limit of Dirac
equation are briefly studied.

\end{abstract}

\maketitle

\newpage

\section{Introduction}

When a particle is exposed to a strong potential field, the relativistic
effect must be considered which gives the correction for nonrelativistic
quantum mechanics. Taking the relativistic effects into account, a spinless
particle in a potential field should be described with the Klein-Gordon (KG)
equation. The solution of the Dirac equation is considered important in
different fields of physics like nuclear and molecular physics [1,2]. Within
the framework of the Dirac equation the spin symmetry arises if the
magnitude of the spherical attractive scalar potential $S$ and repulsive
vector $V$ potential are nearly equal such that $S\sim V$ in the nuclei (%
\textit{i.e.}, when the difference potential $\Delta =V-S=C_{-}=C_{\Delta },$
with $C_{\Delta }$ is an arbitrary constant$).$ However, the pseudospin
(pspin) symmetry occurs if $\ S\sim -V$ are nearly equal (\textit{i.e.},
when the sum potential $\Sigma =V+S=C_{+}=C_{\Sigma },$ with $C_{\Sigma }$
is an arbitrary constant$)$ [3]$.$ The spin symmetry is relevant for mesons
[4]. The pspin symmetry concept has been applied to many systems in nuclear
physics and related areas [3-7] and used to explain features of deformed
nuclei [8], the super-deformation [9] and to establish an effective nuclear
shell-model scheme [5,6,10]. Recently, the spin and pspin symmetries have
been widely applied on several physical potentials by many authors [11-26].
For example, the Dirac equation has been solved for the deformed generalized
Poschl-Teller (PT) potential [27], modified PT potential [28,29],
Manning-Rosen (MR) potential [30], well potential [31], modified Rosen-Morse
(RM) potential [32] and class of potentials including harmonic oscillator,
Morse, Hulth\'{e}n, Scarf' Eckart, MR, Trigonometric RM potentials and
others [33] in the framework of the approximation to the spin-orbit
centrifugal term using the proper quantization rule, algebraic methods,
Ladder operators and su(2) algebra.

The exact solutions of the Dirac equation for the exponential-type
potentials are possible only for the $s$-wave ($\kappa =\pm 1$ case) when
the spin-orbit coupling term will get suppressed [34]. However, for $l$%
-states an approximation scheme has to be used to deal with the spin-orbit
centrifugal $\kappa (\kappa +1)/r^{2}$ (pseudo-centrifugal, $\kappa (\kappa
-1)/r^{2}$) term. In this direction, many works have been done to solve the
Dirac equation with large number of potentials to obtain the energy equation
and the two-component spinor wave functions [35-42]. It has been concluded
that the values of energy spectra may not depend on the spinor structure of
the particle [43], i.e., whether one has a spin-$1/2$ or a spin-$0$
particle. Also, a spin-$1/2$ or a spin-$0$ particle with the same mass and
subject to the same scalar $S$ and vector $V$ potentials of equal magnitude,
i.e., $S=\pm V$ ($\Delta =\Sigma =0$ or $C_{\pm }=0$)$,$ will have the same
energy spectrum (isospectrality), including both bound and scattering states
[43]. It has been shown that for massless particles (or ultrarelativistic
particles) the spin- and pspin spectra of Dirac particles are the same for
the harmonic oscillator potentials [44].

Recently, we obtained the spin symmetric and pspin bound state solutions of
the Dirac equation with the standard RM well potential model [20,45]:%
\begin{equation}
V(r)=-V_{1}\sec h^{2}\alpha r+V_{2}\tanh \alpha r,
\end{equation}%
where the coupling constants $V_{1}$ and $V_{2}$ denote the depth of
the potential and $\alpha $ is the range of the potential that has
an inverse of length dimension. We use the computer software MATLAB
and plot the potential (1) for three different set of parameters
$V_{1}$ and $V_{2}.$ It is plotted in Fig. {1}.

The aim of the present paper is to extend the $s$-wave solutions by solving
the Dirac equation with some physical potentials given in Ref. [34] in the
framework of the Nikiforov-Uvarov (NU) method [46-50] by taking an
approximation to deal with the centrifugal (pseudo-centrifugal) potential
term [20,51]. The approximation scheme used to deal with the spin-orbit
centrifugal barrier $\kappa (\kappa +1)/r^{2}$ holds for values of
spin-orbit coupling quantum number $\kappa $ that are not large and
vibrations of the small amplitude [51]. In the presence of spin symmetry $%
S\sim V$ and pspin symmetry $S\sim -V$, we calculate bound state energy
eigenvalues and their corresponding upper and lower spinor wave functions.
We also show that the spin and pspin symmetry Dirac solutions when $\Delta
=C_{-}$ and $\Sigma =C_{+}$ can be reduced to the exact spin symmetry and
pspin symmetry limitation $\Delta =0$ and $\Sigma =0,$ respectively. \ These
are found to be identical to the KG solution for the $V=\pm S$ cases$.$
Furthermore, the bound state solutions of the Schr\"{o}dinger equation are
also obtained from the nonrelativistic limit of the Dirac equation if an
appropriate mapping of parameters is used.

The paper is organized as follows: Section 2 is mainly devoted to the basic
spin and pspin Dirac equation. In sect. 3, the approximate analytical bound
state solutions of the ($3+1$)-dimensional Dirac equation with the
reflectionless-type and the RM potentials are obtained in the presence of
the spin and pspin limits using a parametric generalization of the NU
method. In sect. 4, special cases like the $s$-wave $\kappa =\pm 1$ ($l=%
\widetilde{l}=0$) and nonrelativistic limit are studied. Section 5 gives the
relevant conclusion.

\section{Basic Spin and Pspin Dirac equations}

The Dirac equation for fermionic massive spin-$1/2$ particles subject to
vector and scalar potentials is [1]
\begin{equation}
\left[ c\mathbf{\alpha }\cdot \mathbf{p+\beta }\left( Mc^{2}+S(r)\right)
+V(r)-E\right] \psi _{n\kappa }(\mathbf{r})=0,\text{ }\psi _{n\kappa }(%
\mathbf{r})=\psi (r,\theta ,\phi ),
\end{equation}%
where $E$ is the binding relativistic energy of the system, $M$ is the mass
of a particle, $\mathbf{p}=-i\hbar \mathbf{\nabla }$ is the momentum
operator, and $\mathbf{\alpha }$ and $\mathbf{\beta }$ are $4\times 4$ Dirac
matrices [3,12,20]. The spinor wave functions take the form%
\begin{equation}
\psi _{n\kappa }(\mathbf{r})=\frac{1}{r}\left(
\begin{array}{c}
F_{n\kappa }(r)Y_{jm}^{l}(\theta ,\phi ) \\
iG_{n\kappa }(r)Y_{jm}^{\widetilde{l}}(\theta ,\phi )%
\end{array}%
\right) ,
\end{equation}%
where $F_{n\kappa }(r)$ and $G_{n\kappa }(r)$ are the radial wave functions
of the upper- and lower-spinor components, respectively, and $%
Y_{jm}^{l}(\theta ,\phi )$ and $Y_{jm}^{\widetilde{l}}(\theta ,\phi )$ are
the spherical harmonic functions coupled to the total angular momentum $j$
and it's projection $m$ on the $z$ axis.

In the presence of spin symmetry ( i.e., $\Delta =C_{-}=C_{\Delta }$), one
obtains a second-order differential equation for the upper-spinor component
[12,20,52,53]:
\begin{equation}
F_{n\kappa }^{\prime \prime }(r)-\left( \frac{\kappa \left( \kappa +1\right)
}{r^{2}}+A_{s}^{2}+B_{s}\Sigma \right) F_{n\kappa }(r)=0,
\end{equation}%
where
\begin{equation}
A_{s}^{2}=\frac{1}{\hbar ^{2}c^{2}}\left[ M^{2}c^{4}-E_{n\kappa }^{2}-\left(
Mc^{2}-E_{n\kappa }\right) C_{-}\right] ,\text{ }B_{s}=\frac{1}{\hbar
^{2}c^{2}}\left( Mc^{2}+E_{n\kappa }-C_{-}\right) ,
\end{equation}%
and $\kappa \left( \kappa +1\right) =l\left( l+1\right) ,$ $\kappa =l$ for $%
\kappa <0$ and $\kappa =-\left( l+1\right) $ for $\kappa >0.$ Further, the
lower-spinor component can be obtained as
\begin{equation}
G_{n\kappa }(r)=\frac{1}{Mc^{2}+E_{n\kappa }-C_{-}}\left( \frac{d}{dr}+\frac{%
\kappa }{r}\right) F_{n\kappa }(r),
\end{equation}%
where $E_{n\kappa }\neq -Mc^{2}$ when $C_{-}=C_{\Delta }=0$ (exact spin
symmetric case). It means that only positive energy spectrum is permitted.

Overmore, in the presence of pspin symmetry ( i.e., $\Sigma =C_{+}=C_{\Sigma
}$), one obtains a second-order differential equation for the lower-spinor
component,%
\begin{equation}
G_{n\kappa }^{\prime \prime }(r)-\left( \frac{\kappa \left( \kappa -1\right)
}{r^{2}}+A_{ps}^{2}-B_{ps}\Delta \right) G_{n\kappa }(r)=0,
\end{equation}%
where
\begin{equation}
A_{ps}^{2}=\frac{1}{\hbar ^{2}c^{2}}\left[ M^{2}c^{4}-E_{n\kappa
}^{2}+\left( Mc^{2}+E_{n\kappa }\right) C_{+}\right] ,\text{ }B_{ps}=\frac{1%
}{\hbar ^{2}c^{2}}\left( Mc^{2}-E_{n\kappa }+C_{+}\right) .
\end{equation}%
The upper-spinor component $F_{n\kappa }(r)$ can be obtained by means of
\begin{equation}
F_{n\kappa }(r)=\frac{1}{Mc^{2}-E_{n\kappa }+C_{+}}\left( \frac{d}{dr}-\frac{%
\kappa }{r}\right) G_{n\kappa }(r),
\end{equation}%
where $E_{n\kappa }\neq Mc^{2}$ when $C_{+}=C_{\Sigma }=0$ (exact pspin
symmetric case). It means that only negative energy spectrum is allowed for
this case. From the above equations, the energy eigenvalues depend on the
quantum numbers $n$ and $\kappa $, and also the pseudo-orbital angular
quantum number $\widetilde{l}$ according to $\kappa (\kappa -1)=\widetilde{l}%
(\widetilde{l}+1),$ which implies that $j=\widetilde{l}\pm 1/2$ are
degenerate for $\widetilde{l}\neq 0.$ The quantum condition for bound states
demands the finiteness of the solution at infinity and at the origin points.

It is known that Eqs. (4) and (7) can be solved exactly only for the case of
$\kappa =-1$ ($l=0$) and $\kappa =1$ ($\widetilde{l}=0$), respectively when
the spin-orbit coupling centrifugal and pseudo-centrifugal terms will get
suppressed. In the case of nonzero $l$ or $\widetilde{l}$ values, we can use
the approximation scheme to deal with the spin-orbit centrifugal
(pseudo-centrifugal) term when $\kappa $ is not large and when vibrations of
the small amplitude near the minimum point $r=r_{e}$ [17,51]
\begin{equation}
\frac{1}{r^{2}}\approx \frac{1}{r_{e}^{2}}\left[ D_{0}+D_{1}\frac{-\exp
(-2\alpha r)}{1+\exp (-2\alpha r)}+D_{2}\left( \frac{-\exp (-2\alpha r)}{%
1+\exp (-2\alpha r)}\right) ^{2}\right] ,
\end{equation}%
where $D_{i}$ is the parameter of coefficients ($i=1,2,3$) given by
\begin{subequations}
\begin{equation}
D_{0}=1-\left( \frac{1+\exp (-2\alpha r_{e})}{2\alpha r_{e}}\right)
^{2}\left( \frac{8\alpha r_{e}}{1+\exp (-2\alpha r_{e})}-\left( 3+2\alpha
r_{e}\right) \right) ,
\end{equation}%
\begin{equation}
D_{1}=-2\left( \exp (2\alpha r_{e})+1\right) \left[ 3\left( \frac{1+\exp
(-2\alpha r_{e})}{2\alpha r_{e}}\right) -\left( 3+2\alpha r_{e}\right)
\left( \frac{1+\exp (-2\alpha r_{e})}{2\alpha r_{e}}\right) \right] ,
\end{equation}%
\begin{equation}
D_{2}=\left( \exp (2\alpha r_{e})+1\right) ^{2}\left( \frac{1+\exp (-2\alpha
r_{e})}{2\alpha r_{e}}\right) ^{2}\left( 3+2\alpha r_{e}-\frac{4\alpha r_{e}%
}{1+\exp (-2\alpha r_{e})}\right) ,
\end{equation}%
and higher order terms are neglected.

\subsection{Spin symmetric solution}

We take the sum potential in Eq. (4) in the form of standard RM well
potential (1), i.e.,
\end{subequations}
\begin{equation}
\Sigma =V(r)=-4V_{1}\frac{\exp (-2\alpha r)}{\left( 1+\exp (-2\alpha
r)\right) ^{2}}+V_{2}\left( \frac{1-\exp (-2\alpha r)}{1+\exp (-2\alpha r)}%
\right) .
\end{equation}%
Upon introducing the new variable $z(r)=\exp (-2\alpha r)$ and substituting
the above sum potential into Eq. (4) which then can be cast into the form%
\begin{equation}
F_{n\kappa }^{\prime \prime }(z)+\frac{\left( 1+z\right) }{z\left(
1+z\right) }F_{n\kappa }^{\prime }(z)+\frac{\left(
-a_{2}z^{2}+a_{1}z-a_{0}^{2}\right) }{z^{2}\left( 1+z\right) ^{2}}F_{n\kappa
}(z)=0,
\end{equation}%
where at boundaries we require that $F_{n\kappa }(0)=F_{n\kappa }(-1)=0$ and
the parameters $a_{i}$ ($i=0,1,2$) take the forms:%
\begin{equation*}
a_{0}=\frac{1}{2\alpha }\sqrt{\frac{\kappa \left( \kappa +1\right) }{%
r_{e}^{2}}D_{0}+B_{s}V_{2}+A_{s}^{2}}>0,
\end{equation*}%
\begin{equation*}
a_{1}=\frac{1}{4\alpha ^{2}}\left( \frac{\kappa \left( \kappa +1\right) }{%
r_{e}^{2}}\left( D_{1}-2D_{0}\right) +4B_{s}V_{1}-2A_{s}^{2}\right) ,
\end{equation*}%
\begin{equation}
a_{2}=\frac{1}{4\alpha ^{2}}\left( \frac{\kappa \left( \kappa +1\right) }{%
r_{e}^{2}}\left( D_{0}-D_{1}+D_{2}\right) +A_{s}^{2}-B_{s}V_{2}\right) .
\end{equation}%
We begin the application of the NU method [46-50] by comparing Eq. (13) with
the hypergeometric differential equation%
\begin{equation}
\psi _{n}^{\prime \prime }(r)+\frac{\widetilde{\tau }(r)}{\sigma (r)}\psi
_{n}^{\prime }(r)+\frac{\widetilde{\sigma }(r)}{\sigma ^{2}(r)}\psi
_{n}(r)=0,\text{ }
\end{equation}%
where%
\begin{equation}
\psi _{n}(r)=\phi (r)y_{n}(r),
\end{equation}%
to identify the parameters,
\begin{equation}
\widetilde{\tau }(z)=1+z,\text{\ }\sigma (z)=z\left( 1+z\right) ,\text{\ }%
\widetilde{\sigma }(z)=-a_{2}z^{2}+a_{1}z-a_{0}^{2},
\end{equation}%
and further calculate the function $\pi (z)$ as
\begin{equation*}
\pi (z)=\frac{1}{2}\left[ \sigma ^{\prime }(r)-\widetilde{\tau }(r)\right]
\pm \sqrt{\frac{1}{4}\left[ \sigma ^{\prime }(r)-\widetilde{\tau }(r)\right]
^{2}-\widetilde{\sigma }(r)+k\sigma (r)}
\end{equation*}%
\begin{equation}
=\frac{z}{2}\pm \frac{1}{2}\sqrt{\left[ 1+4\left( a_{2}+k\right) \right]
z^{2}+4\left( k-a_{1}\right) z+4a_{0}^{2}}.
\end{equation}%
Now we also seek for a physical value of $k$ that makes the discriminant of
the expression under square root in Eq. (18) to be zero, that is%
\begin{equation*}
k=a_{1}+2a_{0}^{2}\pm 2a_{0}q,
\end{equation*}%
\begin{equation}
q=\sqrt{1+\frac{\kappa \left( \kappa +1\right) D_{2}}{\alpha ^{2}r_{e}^{2}}+%
\frac{4V_{1}B_{s}}{\alpha ^{2}}}.
\end{equation}%
Upon the substitution of the value of $k$ into Eq. (18), we obtain the
following convenient solutions:%
\begin{equation}
\pi (z)=-a_{0}-\frac{1}{2}\left( 2a_{0}+q-1\right) z,
\end{equation}%
and%
\begin{equation}
k=a_{1}+2a_{0}^{2}+a_{0}q.
\end{equation}%
With regard to Eqs. (17) and (20), we can calculate the function $\tau (z)=%
\widetilde{\tau }(z)+2\pi (z),$ taking into consideration the bound state
condition which has to be established when $\tau ^{\prime }(z)<0,$ as%
\begin{equation}
\tau (z)=1-2a_{0}-\left( 2a_{0}+q-2\right) z,\text{ }\tau ^{\prime
}(z)=-\left( 2a_{0}+q-2\right) <0,
\end{equation}%
where prime denotes the derivative with respect to $z.$ According to the
method [46-50], in order to find the energy equation from which one
calculates the energy eigenvalues, we need to find the values of the
parameters: $\lambdabar =k+\pi ^{\prime }(s)$ and $\lambdabar =\lambdabar
_{n}=-n\tau ^{\prime }(s)-\frac{1}{2}n\left( n-1\right) \sigma ^{\prime
\prime }(s),\ n=0,1,2,\cdots ,$ as%
\begin{equation}
\lambdabar =\frac{1}{2}+a_{1}+2a_{0}^{2}-a_{0}+\left( a_{0}-\frac{1}{2}%
\right) q,
\end{equation}%
and%
\begin{equation}
\lambdabar _{n}=-n^{2}-n+n\left( 2a_{0}+q\right) ,\text{ }n=0,1,2,\cdots ,
\end{equation}%
respectively. Using the relation $\lambdabar =\lambdabar _{n}$ and the
definitions of variables in Eqs. (14) and (19), we obtain the transcendental
energy equation of relativistic spin-$1/2$ particles in the presence of
vector and scalar potential,%
\begin{equation*}
\frac{1}{\hbar ^{2}c^{2}}\left[ M^{2}c^{4}-E_{n\kappa }^{2}-\left(
Mc^{2}-E_{n\kappa }\right) C_{-}\right] =-\frac{\kappa \left( \kappa
+1\right) D_{0}}{r_{e}^{2}}-B_{s}V_{2}
\end{equation*}%
\begin{equation}
+\alpha ^{2}\left[ n+\frac{1}{2}-\frac{q}{2}+\frac{\kappa \left( \kappa
+1\right) \left( D_{1}-D_{2}\right) /r_{e}^{2}+2B_{s}V_{2}}{4\alpha
^{2}\left( n+\frac{1}{2}-\frac{q}{2}\right) }\right] ^{2},
\end{equation}%
Furthermore, in the exact spin symmetric case $($i.e., $V=S,$ $\Delta =0,$ $%
C_{-}\rightarrow 0)$, we obtain the arbirary $l$-wave energy equation in the
KG theory with equally mixed RM-type potentials (in units of $\hbar =c=1$),%
\begin{equation*}
M^{2}-E_{nl}^{2}=-\frac{l\left( l+1\right) D_{0}}{r_{e}^{2}}-\left(
E_{nl}+M\right) V_{2}
\end{equation*}%
\begin{equation}
+\alpha ^{2}\left[ n+\delta +\frac{l\left( l+1\right) \left(
D_{1}-D_{2}\right) /r_{e}^{2}+2\left( E_{nl}+M\right) V_{2}}{4\alpha
^{2}\left( n+\delta \right) }\right] ^{2},
\end{equation}%
with%
\begin{equation}
\delta =\frac{1}{2}-\frac{1}{2}\sqrt{1+\frac{l\left( l+1\right) D_{2}}{%
\alpha ^{2}r_{e}^{2}}+\frac{4}{\alpha ^{2}}\left( E_{nl}+M\right) V_{1}},
\end{equation}%
where the quantum number $n=0,1,2,\cdots ,$ and the orbital quantum number $%
l=0,1,2,\cdots .$ Actually, the above expression resembles Eq. (13) reported
in [54] when $l=0$ ($s$-wave case). Now, we are going to find the
corresponding wave functions for the present potential model. Firstly, we
calculate the weight function defined as%
\begin{equation}
\rho (z)=\frac{1}{\sigma (z)}\exp \left( \int \frac{\tau (z)}{\sigma (z)}%
dz\right) =z^{-2a_{0}}\left( 1+z\right) ^{-q},
\end{equation}%
and the first part of the wave function in Eq. (16) as%
\begin{equation}
\phi (z)=\exp \left( \int \frac{\pi (z)}{\sigma (z)}dz\right)
=z^{-a_{0}}\left( 1+z\right) ^{\frac{1}{2}(1-q)}.
\end{equation}%
Hence, the second part of the wave function in relation (16) can be obtained
by means of the so called Rodrigues representation
\begin{equation*}
y_{n}(z)=\frac{K_{n}}{\rho (r)}\frac{d^{n}}{dr^{n}}\left[ \sigma ^{n}(r)\rho
(r)\right] =K_{n}z^{2a_{0}}\left( 1+z\right) ^{q}\frac{d^{n}}{dz^{n}}\left[
z^{n-2a_{0}}\left( 1+z\right) ^{n-q}\right]
\end{equation*}%
\begin{equation}
\sim P_{n}^{\left( -2a_{0},-q\right) }(1+2z),\text{ }z\in \lbrack 0,1],
\end{equation}%
where the Jacobi polynomials $P_{n}^{\left( \mu ,\nu \right) }(x)$
are defined for $Re$($\nu )>-1$ and $\Re$($\mu )>-1$ for the
argument $x\in \left[ -1,+1\right] $ and $K_{n}$ is the
normalization constant$.$ By using $F_{n\kappa }(z)=\phi
(z)y_{n}(z),$ in this way we may
write the upper-spinor wave function in the fashion%
\begin{equation*}
F_{n\kappa }(r)=K_{n\kappa }\left( \exp (-2\alpha r)\right) ^{-a_{0}}\left(
1+\exp (-2\alpha r)\right) ^{\frac{1}{2}(1-q)}P_{n}^{\left(
-2a_{0},-q\right) }(1+2\exp (-2\alpha r))
\end{equation*}%
\begin{equation}
=N_{n\kappa }\left( \exp (-2\alpha r)\right) ^{-a_{0}}\left( 1+\exp
(-2\alpha r)\right) ^{\frac{1}{2}(1-q)}%
\begin{array}{c}
_{2}F_{1}%
\end{array}%
\left( -n,n+1-2a_{0}-q;-2a_{0}+1;\exp (-2\alpha r)\right) ,
\end{equation}%
where $a_{0}>0,$ $q>-1.$ The calculated normalization constants $K_{n\kappa }
$ for the upper-spinor component are%
\begin{equation}
N_{n\kappa }=\left[ \frac{\Gamma (-q+2)\Gamma (-2a_{0}+1)}{2\alpha \Gamma (n)%
}\sum\limits_{m=0}^{\infty }\frac{(-1)^{m}\left( n-2a_{0}+1-q\right)
_{m}\Gamma (n+m)}{m!\left( m-2a_{0}\right) !\Gamma \left(
m-2a_{0}-q+2\right) }f_{n\kappa }\right] ^{-1/2}\text{ ,}
\end{equation}%
with
\begin{equation}
f_{n\kappa }=%
\begin{array}{c}
_{3}F_{2}%
\end{array}%
\left( -2a_{0}+m,-n,n+1-2a_{0}-q;m-2a_{0}-q+2;1-2a_{0};1\right) .
\end{equation}%
In addition, the corresponding lower component $G_{n\kappa }(r)$ can be
obtained as follows
\begin{equation*}
G_{n\kappa }(r)=c_{n\kappa }\frac{\left( \exp (-2\alpha r)\right)
^{-a_{0}}(1+\exp (-2\alpha r))^{\frac{1}{2}(1-q)}}{\left( Mc^{2}+E_{n\kappa
}-C_{-}\right) }\left[ -2\alpha a_{0}-\frac{\alpha \left( 1-q\right) \exp
(-2\alpha r)}{\left( 1+\exp (-2\alpha r)\right) }+\frac{\kappa }{r}\right]
\end{equation*}%
\begin{equation*}
\times
\begin{array}{c}
_{2}F_{1}%
\end{array}%
\left( -n,n-a_{0}-q+1;-2a_{0}+1;\exp (-2\alpha r)\right)
\end{equation*}%
\begin{equation*}
+c_{n\kappa }\left[ \frac{2\alpha n\left[ n-2a_{0}-q+1\right] \left( \exp
(-2\alpha r)\right) ^{-a_{0}+1}\left( 1+\exp (-2\alpha r)\right) ^{\frac{1}{2%
}(1-q)}}{\left( 2a_{0}+1\right) \left( Mc^{2}+E_{n\kappa }-C_{-}\right) }%
\right]
\end{equation*}%
\begin{equation}
\times
\begin{array}{c}
_{2}F_{1}%
\end{array}%
\left( -n+1;n-2a_{0}-q+2;-2a_{0}+2;\exp (-2\alpha r)\right) ,\text{ }a_{0}>0,
\end{equation}%
where $E_{n\kappa }\neq -Mc^{2}$ for exact spin symmetry. Here, note that
the hypergeometric series $%
\begin{array}{c}
_{2}F_{1}%
\end{array}%
\left( -n,n-2a_{0}-q+1;-2a_{0}+1;\exp (-2\alpha r)\right) $ terminates for $%
n=0$ and thus converge for all values of real parameters $q>0$ and $a_{0}>0.$

\subsection{Pspin symmetric solution}

In the same way as before, this time taking the difference potential in Eq.
(7) as%
\begin{equation}
\Delta =V(r)=-4V_{1}\frac{\exp (-2\alpha r)}{\left( 1+\exp (-2\alpha
r)\right) ^{2}}+V_{2}\left( \frac{1-\exp (-2\alpha r)}{1+\exp (-2\alpha r)}%
\right) ,
\end{equation}%
and in terms of new variable $z(r)=\exp (-2\alpha r),$ leads us to obtain a
Schr\"{o}dinger-like equation for the lower-spinor component $G_{n\kappa
}(r),$%
\begin{equation}
G_{n\kappa }^{\prime \prime }(z)+\frac{\left( 1+z\right) }{z\left(
1+z\right) }G_{n\kappa }^{\prime }(z)+\frac{\left(
-b_{2}z^{2}+b_{1}z-b_{0}^{2}\right) }{z^{2}\left( 1+z\right) ^{2}}G_{n\kappa
}(z)=0,
\end{equation}%
where the parameters $b_{j}$ ($j=0,1,2$) are defined by%
\begin{equation*}
b_{0}=\frac{1}{2\alpha }\sqrt{\frac{\kappa \left( \kappa -1\right) }{%
r_{e}^{2}}D_{0}+B_{ps}V_{2}+A_{ps}^{2}}>0,
\end{equation*}%
\begin{equation*}
b_{1}=\frac{1}{4\alpha ^{2}}\left( \frac{\kappa \left( \kappa -1\right) }{%
r_{e}^{2}}\left( D_{1}-2D_{0}\right) +4B_{ps}V_{1}-2A_{ps}^{2}\right) ,
\end{equation*}%
\begin{equation}
b_{2}=\frac{1}{4\alpha ^{2}}\left( \frac{\kappa \left( \kappa -1\right) }{%
r_{e}^{2}}\left( D_{0}-D_{1}+D_{2}\right) +A_{ps}^{2}-B_{ps}V_{2}\right) .
\end{equation}%
To avoid repetition in the solution of Eq. (36), a first inspection for the
relationship between the present set of parameters $(b_{0},b_{1},b_{2})$ and
the previous set $(a_{0},a_{1},a_{2})$ tells us that the negative energy
solution for pseudospin symmetry such that $\Sigma =C_{+}=C_{\Sigma }$ can
be obtained directly from those of the positive energy solution above for
spin symmetry by performing the changes [12,20]:
\begin{equation}
F_{n\kappa }(r)\leftrightarrow G_{n\kappa }(r),\text{ }V(r)\rightarrow -V(r)%
\text{ (or }V_{1}\rightarrow -V_{1},\text{ }V_{2}\rightarrow -V_{2}\text{ )},%
\text{ }E_{n\kappa }\rightarrow -E_{n\kappa }\text{ and }C_{-}\rightarrow
-C_{+}.
\end{equation}%
Considering the previous results in Eq. (25) and applying the above
transformations, we finally arrive at the pspin symmetric energy equation%
\begin{equation*}
\left[ M^{2}c^{4}-E_{n\kappa }^{2}+\left( Mc^{2}+E_{n\kappa }\right) C_{+}%
\right] =-\frac{\hbar ^{2}c^{2}\kappa \left( \kappa -1\right) D_{0}}{%
r_{e}^{2}}+\left( Mc^{2}-E_{n\kappa }+C_{+}\right) V_{2}
\end{equation*}%
\begin{equation}
+\frac{\hbar ^{2}c^{2}\alpha ^{2}}{4}\left[ 2n+1-p+\frac{\hbar
^{2}c^{2}\kappa \left( \kappa -1\right) \left( D_{1}-D_{2}\right)
/r_{e}^{2}-2\left( Mc^{2}-E_{n\kappa }+C_{+}\right) V_{2}}{\hbar
^{2}c^{2}\alpha ^{2}\alpha ^{2}\left( 2n+1-p\right) }\right] ^{2},
\end{equation}%
where%
\begin{equation}
p=\sqrt{1+\frac{\kappa \left( \kappa -1\right) D_{2}}{\alpha ^{2}r_{e}^{2}}-%
\frac{4V_{1}B_{ps}}{\alpha ^{2}}}.
\end{equation}%
Again, the radial lower-spinor wave function in Eq. (31) becomes%
\begin{equation*}
G_{n\kappa }(r)=d_{n\kappa }\left( \exp (-2\alpha r)\right) ^{-b_{0}}\left(
1+\exp (-2\alpha r)\right) ^{\frac{1}{2}(1-p)}P_{n}^{\left(
-2b_{0},-p\right) }(1+2\exp (-2\alpha r)).
\end{equation*}%
\begin{equation}
=d_{n\kappa }\left( \exp (-2\alpha r)\right) ^{-b_{0}}\left( 1+\exp
(-2\alpha r)\right) ^{\frac{1}{2}(1-p)}%
\begin{array}{c}
_{2}F_{1}%
\end{array}%
\left( -n,n+-2b_{0}-p+1;-2b_{0}+1;\exp (-2\alpha r)\right) ,
\end{equation}%
which satisfies the restriction condition for the bound states, \textit{i.e.}%
, $\ p>0$ and $b_{0}>0$ and the normalization constants is%
\begin{equation}
d_{n\kappa }=\left[ \frac{\Gamma (-p+2)\Gamma (-2b_{0}+1)}{2\alpha \Gamma (n)%
}\sum\limits_{m=0}^{\infty }\frac{(-1)^{m}\left(
n+1-2b_{0}-p)\right) _{m}\Gamma (n+m)}{m!\left( m-2b_{0}\right)
!\Gamma \left( m-2b_{0}-p+2\right) }g_{n\kappa }\right]
^{-1/2}\text{ ,}
\end{equation}%
with
\begin{equation}
g_{n\kappa }=%
\begin{array}{c}
_{3}F_{2}%
\end{array}%
\left( -2b_{0}+m,-n,n+1-2b_{0}-p);m-2b_{0}-p+2;1-2b_{0};1\right) .
\end{equation}

\section{Applications to Some Physical Potential Models}

We adopt the following two physical potential cases that belong to the
general potential model been introduced in Eq. (1).

\subsection{The reflectionless-type potential}

The reflectionless-type potential is the special case of the symmetrical
double-well potential.offered by B\"{u}y\"{u}kk\i l\i \c{c} \textit{et al}
[55] to describe the vibration of polyatomic molecules. This can be achieved
when the coefficient of $\tanh \alpha r$ becomes zero. So, it takes the form%
\begin{equation}
V(r)=-a^{2}\sec h^{2}\alpha r,\text{ }a^{2}=\lambda (\lambda +1)/2,\text{ }%
\lambda =1,2,3,\cdots .
\end{equation}%
The potential is plotted in Fig. 2 for three different values $\lambda =1,2$
and $3.$ It follows that the energy equation in Eq. (25) becomes%
\begin{equation}
M^{2}-E_{n\kappa }^{2}-C_{-}\left( M-E_{n\kappa }\right) =-\frac{\kappa
\left( \kappa +1\right) D_{0}}{r_{e}^{2}}+\frac{\alpha ^{2}}{4}\left[
2n+1-q_{0}+\frac{\kappa \left( \kappa +1\right) \left( D_{1}-D_{2}\right)
/r_{e}^{2}}{\alpha ^{2}\left( 2n+1-q_{0}\right) }\right] ^{2},
\end{equation}%
and the upper-spinor wave functions from Eq. (31) turns to be%
\begin{equation}
F_{n\kappa }(r)=N_{n\kappa }\left( \exp (-2\alpha r)\right) ^{-s_{0}}\left(
1+\exp (-2\alpha r)\right) ^{\frac{1}{2}(1-q_{0})}P_{n}^{\left(
-2s_{0},-q_{0}\right) }(1+2\exp (-2\alpha r)),
\end{equation}%
where%
\begin{equation}
s_{0}=\frac{1}{2\alpha }\sqrt{\frac{\kappa \left( \kappa +1\right) }{%
r_{e}^{2}}D_{0}+A_{s}^{2}}>0,\text{ }q_{0}=\sqrt{1+\frac{\kappa \left(
\kappa +1\right) D_{2}}{\alpha ^{2}r_{e}^{2}}+\frac{4a^{2}B_{s}}{\alpha ^{2}}%
}.
\end{equation}%
It is worth noting that the results given above in Eq. (45) and (46) are
identical to those ones of Ref. [27] for $s$-wave case $(\kappa =-1)$. In
the presence of the pspin case, the energy spectrum becomes%
\begin{equation}
M^{2}-E_{n\kappa }^{2}+C_{+}\left( M+E_{n\kappa }\right) =-\frac{\kappa
\left( \kappa -1\right) D_{0}}{r_{e}^{2}}+\frac{\alpha ^{2}}{4}\left[
2n+1-p_{0}+\frac{\kappa \left( \kappa -1\right) \left( D_{1}-D_{2}\right)
/r_{e}^{2}}{\alpha ^{2}\left( 2n+1-p_{0}\right) }\right] ^{2},
\end{equation}%
and the lower spinor component of pseudospin symmetric wave function%
\begin{equation}
G_{n\kappa }(r)=\widetilde{N}_{n\kappa }\left( \exp (-2\alpha r)\right)
^{-w_{0}}\left( 1+\exp (-2\alpha r)\right) ^{\frac{1}{2}(1-p_{0})}P_{n}^{%
\left( -2w_{0},-p_{0}\right) }(1+2\exp (-2\alpha r)).
\end{equation}%
where%
\begin{equation}
w_{0}=\frac{1}{2\alpha }\sqrt{\frac{\kappa \left( \kappa -1\right) }{%
r_{e}^{2}}D_{0}+A_{ps}^{2}}>0,\text{ }p_{0}=\sqrt{1+\frac{\kappa \left(
\kappa -1\right) D_{2}}{\alpha ^{2}r_{e}^{2}}-\frac{4a^{2}B_{ps}}{\alpha ^{2}%
}}>0,
\end{equation}%
$4a^{2}B_{ps}/\alpha ^{2}\leq 1$ for bound states when $\kappa =1.$

Let us now discuss the non-relativistic limit of the energy eigenvalues and
wave functions of our solution. If we take $C_{-}=0$ ($\Delta =0)$ and
consider the transformations $E_{n\kappa }+M\simeq 2\mu $ and $E_{n\kappa
}-M\simeq E_{nl}$ [52,53], we would have the following expression for the
energy equation (45) and wave functions (46) (in $\hbar =c=1$)%
\begin{equation}
E_{nl}=\frac{l\left( l+1\right) D_{0}}{2\mu r_{e}^{2}}-\frac{\alpha ^{2}}{%
2\mu }\left[ n+\frac{1}{2}-\frac{1}{2}q_{0}+\frac{\hbar ^{2}l\left(
l+1\right) \left( D_{1}-D_{2}\right) }{4r_{e}^{2}\alpha ^{2}\left( n+\frac{1%
}{2}-\frac{1}{2}q_{0}\right) }\right] ^{2},
\end{equation}%
and the wave functions:%
\begin{equation}
R_{nl}(r)=N_{nl}\left( \exp (-2\alpha r)\right) ^{-s_{0}}\left( 1+\exp
(-2\alpha r)\right) ^{\frac{1}{2}(1-q_{0})}P_{n}^{\left(
-2s_{0},-q_{0}\right) }(1+2\exp (-2\alpha r)),
\end{equation}%
with%
\begin{equation}
s_{0}=\frac{1}{2\alpha }\sqrt{\frac{l\left( l+1\right) }{r_{e}^{2}}D_{0}-%
\frac{2\mu }{\hbar ^{2}}E_{nl}}>0,\text{ }q_{0}=\sqrt{1+\frac{l\left(
l+1\right) D_{2}}{\alpha ^{2}r_{e}^{2}}+\frac{8\mu a^{2}}{\alpha ^{2}\hbar
^{2}}},
\end{equation}%
where $E_{nl}<\frac{l\left( l+1\right) }{2\mu r_{e}^{2}}D_{0}$ is a
condition for bound state solutions. 2.

To conclude, it is necessary to mention that the reflectionless-type
potential here reminds one of the modified PT potential in the
one-dimensional (1D) case [29]. However, for the present case it is in the
three-dimensional (3D) case. Thus, the original symmetry is broken. The
energy levels could be obtained readily.

\subsection{The Rosen-Morse potential}

The standard RM potential was given by Rosen and Morse in Ref. [45] useful
to describe interatomic interaction of the linear molecules and helpful for
discussing polyatomic vibrational energies. As example of its application to
the to the vibrational states of the $NH_{3}$ molecule. This can be achieved
when%
\begin{equation}
V(r)=-a(a+\alpha )\sec h^{2}\alpha r+2b\tanh \alpha r,
\end{equation}%
where $a$ and $b$ are real dimensionless parameters. In Fig. 3, we plot this
potential for three various sets of parameter values. It follows that from
Eqs. (25) and (31), the spin symmetry energy spectrum for the RM well is%
\begin{equation*}
M^{2}-E_{n\kappa }^{2}-C_{-}\left( M-E_{n\kappa }\right) =-\frac{\kappa
\left( \kappa +1\right) D_{0}}{r_{e}^{2}}-2b\left( M+E_{n\kappa
}-C_{-}\right)
\end{equation*}%
\begin{equation}
+\frac{\alpha ^{2}}{4}\left[ 2n+1-q_{1}+\frac{\kappa \left( \kappa +1\right)
\left( D_{1}-D_{2}\right) /r_{e}^{2}+4b\left( M+E_{n\kappa }-C_{-}\right) }{%
\alpha ^{2}\left( 2n+1-q_{1}\right) }\right] ^{2},
\end{equation}%
and the upper spinor component $F_{n\kappa }(r)$ of the wave functions as%
\begin{equation}
F_{n\kappa }(r)=N_{n\kappa }\left( \exp (-2\alpha r)\right) ^{-s_{1}}\left(
1+\exp (-2\alpha r)\right) ^{\frac{1}{2}(1-q_{1})}P_{n}^{\left(
-2s_{1},-q_{1}\right) }(1+2\exp (-2\alpha r)),
\end{equation}%
respectively, where%
\begin{equation}
s_{1}=\frac{1}{2\alpha }\sqrt{\frac{\kappa \left( \kappa +1\right) }{%
r_{e}^{2}}D_{0}+2bB_{s}+A_{s}^{2}}>0,\text{ }q_{1}=\sqrt{1+\frac{\kappa
\left( \kappa +1\right) D_{2}}{\alpha ^{2}r_{e}^{2}}+\frac{4a(a+\alpha )B_{s}%
}{\alpha ^{2}}}.
\end{equation}%
Overmore, in the presence of the pspin symmetry, the energy spectrum for the
RM well is%
\begin{equation*}
M^{2}-E_{n\kappa }^{2}+C_{+}\left( M+E_{n\kappa }\right) =-\frac{\kappa
\left( \kappa -1\right) D_{0}}{r_{e}^{2}}+2b\left( M-E_{n\kappa
}+C_{+}\right)
\end{equation*}%
\begin{equation}
+\frac{\alpha ^{2}}{4}\left[ 2n+1-p_{1}+\frac{\kappa \left( \kappa -1\right)
\left( D_{1}-D_{2}\right) /r_{e}^{2}-4b\left( M-E_{n\kappa }+C_{+}\right) }{%
\alpha ^{2}\left( 2n+1-p_{1}\right) }\right] ^{2},
\end{equation}%
and the lower-spinor wave function is%
\begin{equation}
G_{n\kappa }(r)=d_{n\kappa }\left( \exp (-2\alpha r)\right) ^{-w_{1}}\left(
1+\exp (-2\alpha r)\right) ^{\frac{1}{2}(1-p_{1})}P_{n}^{\left(
-2w_{1},-p_{1}\right) }(1+2\exp (-2\alpha r)),
\end{equation}%
with%
\begin{equation}
w_{1}=\frac{1}{2\alpha }\sqrt{\frac{\kappa \left( \kappa -1\right) }{%
r_{e}^{2}}D_{0}+2bB_{ps}+A_{ps}^{2}}>0,\text{ }p_{1}=\sqrt{1+\frac{\kappa
\left( \kappa -1\right) D_{2}}{\alpha ^{2}r_{e}^{2}}-\frac{4a(a+\alpha
)B_{ps}}{\alpha ^{2}}},
\end{equation}%
where $4a(a+\alpha )B_{ps}/\alpha ^{2}\leq 1$ when $\kappa =1.$ Let us now
discuss the non-relativistic limit of the energy eigenvalues and wave
functions of our solution. If we take $C_{-}=0$ ($\Delta =0)$ and consider
the nonrelativistic limits [52,53], we would have the following expression
for the energy equation (55) and the upper spinor component of the wave
functions (56) (in units $\hbar =c=1$)%
\begin{equation*}
E_{nl}=\frac{l\left( l+1\right) D_{0}}{2\mu r_{e}^{2}}+2b
\end{equation*}%
\begin{equation}
-\frac{\alpha ^{2}}{2\mu }\left[ n+\frac{1}{2}-\frac{1}{2}\sqrt{1+\frac{%
l\left( l+1\right) D_{2}}{\alpha ^{2}r_{e}^{2}}+\frac{8\mu a(a+\alpha )}{%
\alpha ^{2}}}+\frac{l\left( l+1\right) \left( D_{1}-D_{2}\right)
/r_{e}^{2}+8\mu b}{\alpha ^{2}\left( n+\frac{1}{2}-\frac{1}{2}\sqrt{1+\frac{%
l\left( l+1\right) D_{2}}{\alpha ^{2}r_{e}^{2}}+\frac{8\mu a(a+\alpha )}{%
\alpha ^{2}}}\right) }\right] ^{2},
\end{equation}%
and%
\begin{equation}
R_{nl}(r)=N_{nl}\left( \exp (-2\alpha r)\right) ^{-s_{1}}\left( 1+\exp
(-2\alpha r)\right) ^{\frac{1}{2}(1-q_{1})}P_{n}^{\left(
-2s_{1},-q_{1}\right) }(1+2\exp (-2\alpha r)),
\end{equation}%
respectively, where%
\begin{equation}
s_{1}=\frac{1}{2\alpha }\sqrt{\frac{l\left( l+1\right) }{r_{e}^{2}}%
D_{0}+4\mu b-2\mu E_{nl}}>0,\text{ }q_{1}=\sqrt{1+\frac{l\left( l+1\right)
D_{2}}{\alpha ^{2}r_{e}^{2}}+\frac{8\mu a(a+\alpha )}{\alpha ^{2}}}.
\end{equation}%
and $N_{nl}$ is the normalization constant.

To conclude, it is necessary to mention that the RM potential was studied
byusing the proper quantization rule in Ref. [33].

\section{Discussions}

We study two special cases of the energy eigenvalues given by Eqs. (25) and
(39) for the spin and pspin symmetry, respectively.

(I) The $s$-wave spin symmetric case ($\kappa =-1,$ $l=0)$ (in units $\hbar
=c=1$). For the reflectionless-type potential, we have%
\begin{equation}
M^{2}-E_{n,-1}^{2}-\left( M-E_{n,-1}\right) C_{-}=\alpha ^{2}\left[ n+\frac{1%
}{2}-\frac{1}{2}\sqrt{1+\frac{4a^{2}}{\alpha ^{2}}\left(
M+E_{n,-1}-C_{-}\right) }\right] ^{2},
\end{equation}%
and%
\begin{equation*}
F_{n,-1}(r)=c_{n,-1}\left( \exp (-2\alpha r)\right) ^{-A_{-1}/2\alpha
}\left( 1+\exp (-2\alpha r)\right) ^{\frac{1}{2}\left( 1-\sqrt{1+\frac{4a^{2}%
}{\alpha ^{2}}\left( M+E_{n,-1}-C_{-}\right) }\right) }
\end{equation*}%
\begin{equation}
\times P_{n}^{\left( -A_{-1}/\alpha ,-\sqrt{1+\frac{4a^{2}}{\alpha ^{2}}%
\left( M+E_{n,-1}-C_{-}\right) }\right) }(1+2\exp (-2\alpha r))
\end{equation}%
where $A_{-1}^{2}=M^{2}-E_{n,-1}^{2}-\left( M-E_{n,-1}\right) C_{-}.$ The $%
\widetilde{s}$-wave pspin symmetric case ($\kappa =1,\widetilde{l}=0$):%
\begin{equation}
M^{2}-E_{n,1}^{2}+C_{+}\left( M+E_{n,1}\right) =\alpha ^{2}\left[ n+\frac{1}{%
2}-\frac{1}{2}\sqrt{1-\frac{4a^{2}}{\alpha ^{2}}\left(
M-E_{n,1}+C_{+}\right) }\right] ^{2},
\end{equation}%
where $\frac{4a^{2}}{\alpha ^{2}}\left( M-E_{n,1}+C_{+}\right) \leq 1$ and
the lower spinor component of pspin symmetric wave function%
\begin{equation*}
G_{n,1}(r)=d_{n,1}\left( \exp (-2\alpha r)\right) ^{-A_{1}/2\alpha }\left(
1+\exp (-2\alpha r)\right) ^{\frac{1}{2}\left( 1-\sqrt{1-\frac{4a^{2}}{%
\alpha ^{2}}\left( M-E_{n,1}+C_{+}\right) }\right) }
\end{equation*}%
\begin{equation}
\times P_{n}^{\left( -A_{1}/\alpha ,-\sqrt{1-\frac{4a^{2}}{\alpha ^{2}}%
\left( M-E_{n1}+C_{+}\right) }\right) }(1+2\exp (-2\alpha r)).
\end{equation}%
where $A_{1}^{2}=M^{2}-E_{n,1}^{2}+\left( M+E_{n,1}\right) C_{+}.$ For the
RM potential model (spin symmetric case), we have%
\begin{equation*}
M^{2}-E_{n,-1}^{2}-C_{-}\left( M-E_{n,-1}\right) =-2b\left(
M+E_{n,-1}-C_{-}\right)
\end{equation*}%
\begin{equation}
+\alpha ^{2}\left[ n+\frac{1}{2}-\frac{\beta _{-1}}{2}+\frac{b\left(
M+E_{n\kappa }-C_{-}\right) }{\alpha ^{2}\left( n+\frac{1}{2}-\frac{\beta
_{-1}}{2}\right) }\right] ^{2},
\end{equation}%
and the upper spinor component $F_{n\kappa }(r)$ of the wave functions as%
\begin{equation*}
F_{n,-1}(r)=N_{n,-1}\left( \exp (-2\alpha r)\right) ^{-\gamma _{-1/2}}\left(
1+\exp (-2\alpha r)\right) ^{\frac{1}{2}\left( 1-\beta _{-1}\right)
}P_{n}^{\left( -\gamma _{-1},-\beta _{-1}\right) }(1+2\exp (-2\alpha r)),
\end{equation*}%
where $\beta _{-1}=\sqrt{1+\frac{4a(a+\alpha )}{\alpha ^{2}}\left(
M+E_{n,-1}-C_{-}\right) }$ and $\gamma _{-1}=\sqrt{2b\left(
M+E_{n,-1}-C_{-}\right) +A_{-1}^{2}}/2\alpha .$ For the pseudospin case:%
\begin{equation*}
M^{2}-E_{n,1}^{2}+C_{+}\left( M+E_{n,1}\right) =2b\left(
M-E_{n,1}+C_{+}\right)
\end{equation*}%
\begin{equation}
+\alpha ^{2}\left[ \left( n+\frac{1}{2}-\frac{\beta _{1}}{2}\right) -\frac{%
b\left( M-E_{n,1}+C_{+}\right) }{\alpha ^{2}\left( n+\frac{1}{2}-\frac{\beta
_{1}}{2}\right) }\right] ^{2},
\end{equation}%
and the lower-spinor wave functions become%
\begin{equation}
G_{n,1}(r)=d_{n,1}\left( \exp (-2\alpha r)\right) ^{-\gamma _{1}/2}\left(
1+\exp (-2\alpha r)\right) ^{\frac{1}{2}\left( 1-\beta _{1}\right)
}P_{n}^{\left( -\gamma _{1},-\beta _{1}\right) }(1+2\exp (-2\alpha r)),
\end{equation}%
where $\beta _{1}=\sqrt{1-\frac{4a(a+\alpha )}{\alpha ^{2}}\left(
M-E_{n,1}+C_{+}\right) }$ and $\gamma _{1}=\sqrt{2b\left(
M-E_{n,1}+C_{+}\right) +A_{1}^{2}}/\alpha .$

(II) The transformation of the potential (1) into other potential forms. For
a potential $V(x),$ when one makes the transformations: $x\rightarrow -x,$ $%
\alpha \rightarrow i\alpha $ and $V_{2}\rightarrow iV_{2}$ (complex
parameters), then Eq. (1) transforms into a trigonometric Rosen-Morse-type
(tRM) form:%
\begin{equation}
V(x)=-V_{1}\sec ^{2}\alpha x+V_{2}\tan \alpha x,\text{ }\alpha =\frac{\pi }{%
2a},\text{ }x=[0,a],
\end{equation}%
where $\Re(V_{1})>0.$ When $x\rightarrow -x$ and $i\rightarrow -i,$
if the relation $V(-x)=V^{\ast }$ exists, the potential $V(x)$ is
said to be $PT $-symmetric, where $P$ denotes parity operator (space
reflection) and $T$
denotes time reversal (see e.g., [56,57] and the references therein). This $%
PT$-symmetric potential is plotted in Figure 4 for various sets of
parameters $V_{1}$ and $V_{2}.$ Thus the spin-symmetric energy equation ($%
\kappa =-1$) can be obtained from Eqs. (19) and (25) as
\begin{equation*}
M^{2}-E_{n,-1}^{2}-\left( M-E_{n,-1}\right) C_{-}=-\alpha ^{2}\left( n+\frac{%
1}{2}-\frac{1}{2}\sqrt{1-\frac{4V_{1}}{\alpha ^{2}}\left(
M+E_{n,-1}-C_{-}\right) }\right) ^{2}
\end{equation*}%
\begin{equation}
+\left( \frac{V_{2}}{2\alpha }\right) ^{2}\left( \frac{M+E_{n,-1}-C_{-}}{n+%
\frac{1}{2}-\frac{1}{2}\sqrt{1-\frac{4V_{1}}{\alpha ^{2}}\left(
M+E_{n,-1}-C_{-}\right) }}\right) ^{2},
\end{equation}%
where $4V_{1}\left( M+E_{n,-1}-C_{-}\right) \leq \alpha ^{2}.$

\section{Final Comments and Conclusion}

In summary, we have obtained the approximate analytic relations for the
relativistic energy spectra and the corresponding upper and lower spinor
wave functions in the presence of spherical scalar and vector
reflectionless-type and RM potential models under the conditions of the spin
and pspin symmetries. The resulting solutions of the wave functions are
being expressed in terms of the generalized Jacobi polynomials or
hypergeometric functions. Parametric generalization of the NU method is
used. We have further used the recently introduced exponential approximation
to deal with the spin-orbit centrifugal (pseudo-centrifugal) potential term.
The most stringent interesting result is that the present spin (pspin)
symmetric energy spectrum of the Dirac equation is noticed to be the same as
the energy spectrum of the KG solution if $V=\pm S$ (\textit{i.e}., $\Sigma
=\Delta =0,$ $C_{\pm }=0$). We point out a possible remark of this result.
The conditions that originate the spin and pspin symmetries in the Dirac
equation are the same that produce equivalent energy spectra of relativistic
spin-$1/2$ and spin-$0$ particles in the presence of spherical vector and
scalar potentials. Obviously, the relativistic solution can be reduced to
its non-relativistic limit by the choice of appropriate mapping
transformations. Also, in case when spin-orbit quantum number $\kappa =\pm 1$
($l=\widetilde{l}=0$)$,$ the problem can be easily reduced to the $s$-wave
solution. We also find that when we let $x\rightarrow -x,$ $\alpha
\rightarrow i\alpha ,$ $V_{2}\rightarrow iV_{2},$ the RM potential (1) turns
into tRM potential (71) with real energy solution. We hope that, as in the
nonrelativistic case (see, for example [58]), the relativistic model under
consideration can be applied in molecular physics as well as nuclear
physics. We stress that the present results should be useful in studying the
rotation-vibration energy spectrum of low vibrational molecules of small
amplitude and spin-orbit quantum nnmber $\kappa $ that are not large [51,58].

\acknowledgments The partial support provided by the Scientific and
Technological Research Council of Turkey (T\"{U}B\.{I}TAK) is highly
appreciated. The authors wish to thank Professor Shi-Hai Dong for his
invaluable suggestions and comments

\newpage\

\ {\normalsize 
}

\newpage

\begin{figure}[htbp]
\centering
\includegraphics[height=4.5in, width=7in,
angle=0]{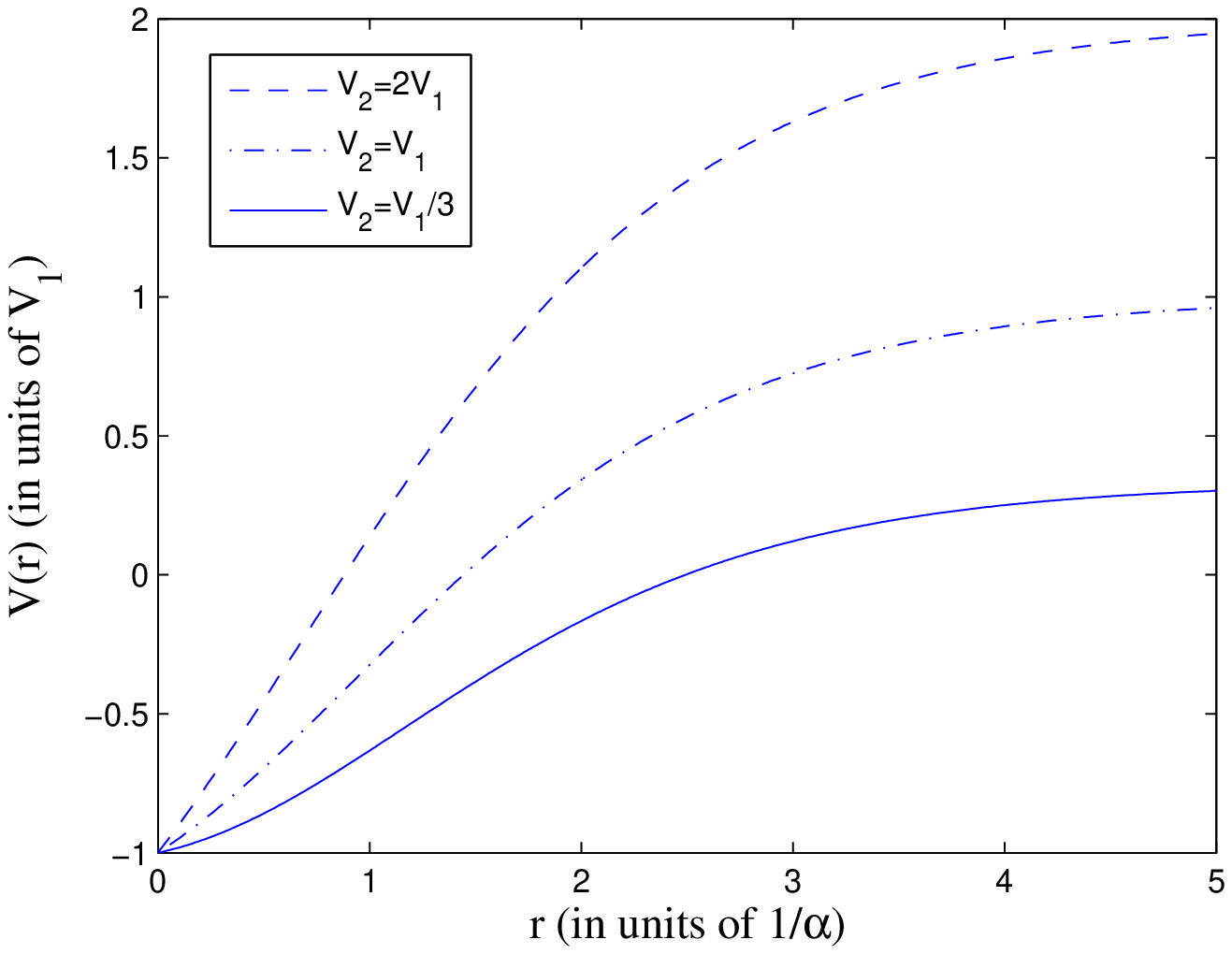} \caption{A plot of the Rosen-Morse-type
potential (1) for three different cases $V_{2}=2V_{1},$ $V_{2}=V_{1}$ and $%
V_{2}=V_{1}/3.$}
\end{figure}

\newpage

\begin{figure}[htbp]
\centering
\includegraphics[height=4.5in, width=7in,
angle=0]{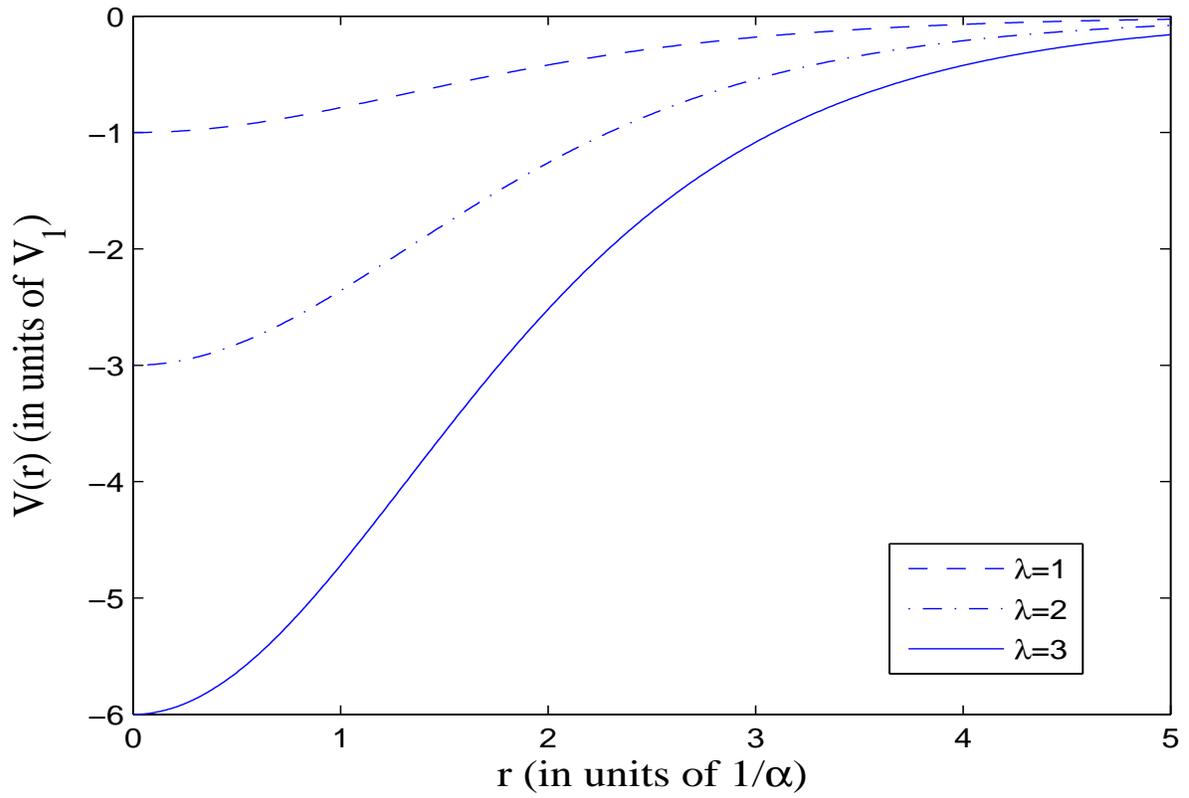} \caption{A plot of the reflectionless potential (44) for three various values $%
\protect\lambda =1,2,$ and $3.$}
\end{figure}

\newpage

\begin{figure}[htbp]
\centering
\includegraphics[height=4.5in, width=7in,
angle=0]{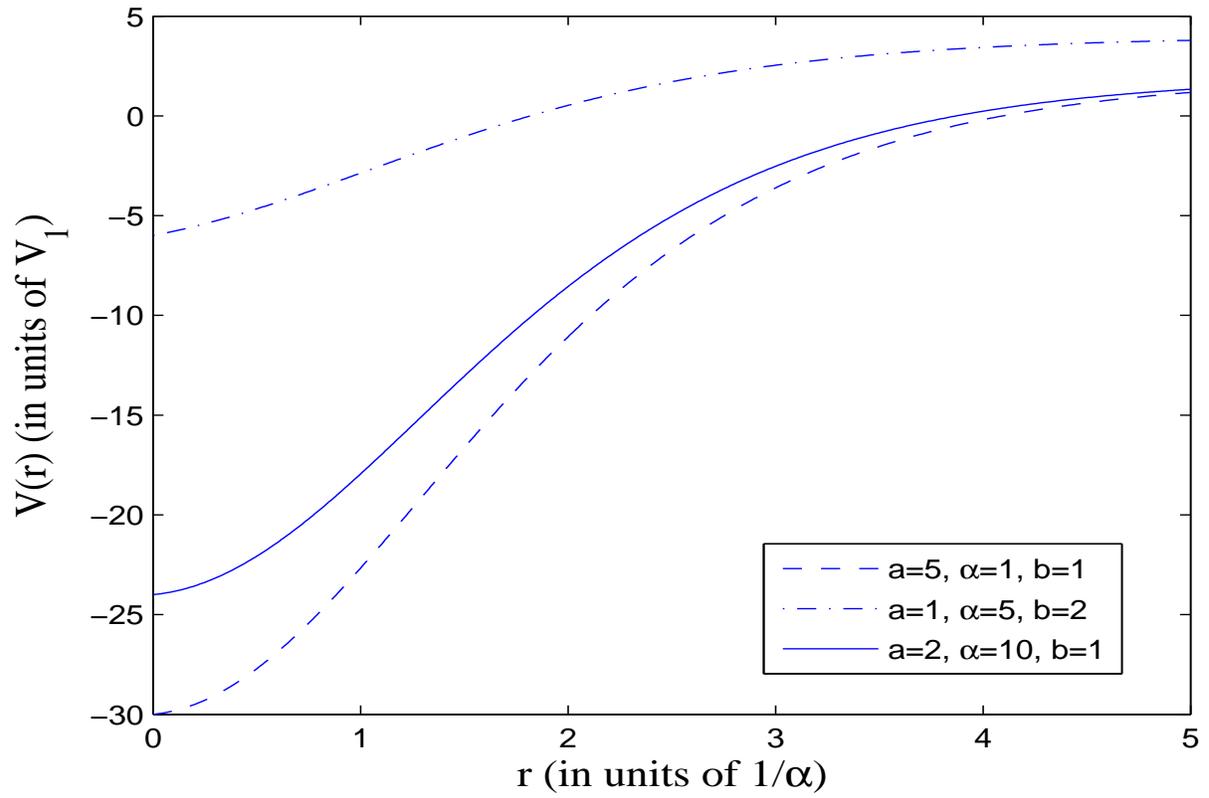} \caption{A plot of the Rosen-Morse potential
(54) for three
different sets of parameters (i) $a=5,$ $\protect\alpha =1,$ $b=1,$ (ii) $%
a=1,$ $\protect\alpha =5,$ $b=2,$ and (iii) $a=2,$ $\protect\alpha =10,$ $%
b=1.$}
\end{figure}

\newpage

\begin{figure}[htbp]
\centering
\includegraphics[height=4.5in, width=7in,
angle=0]{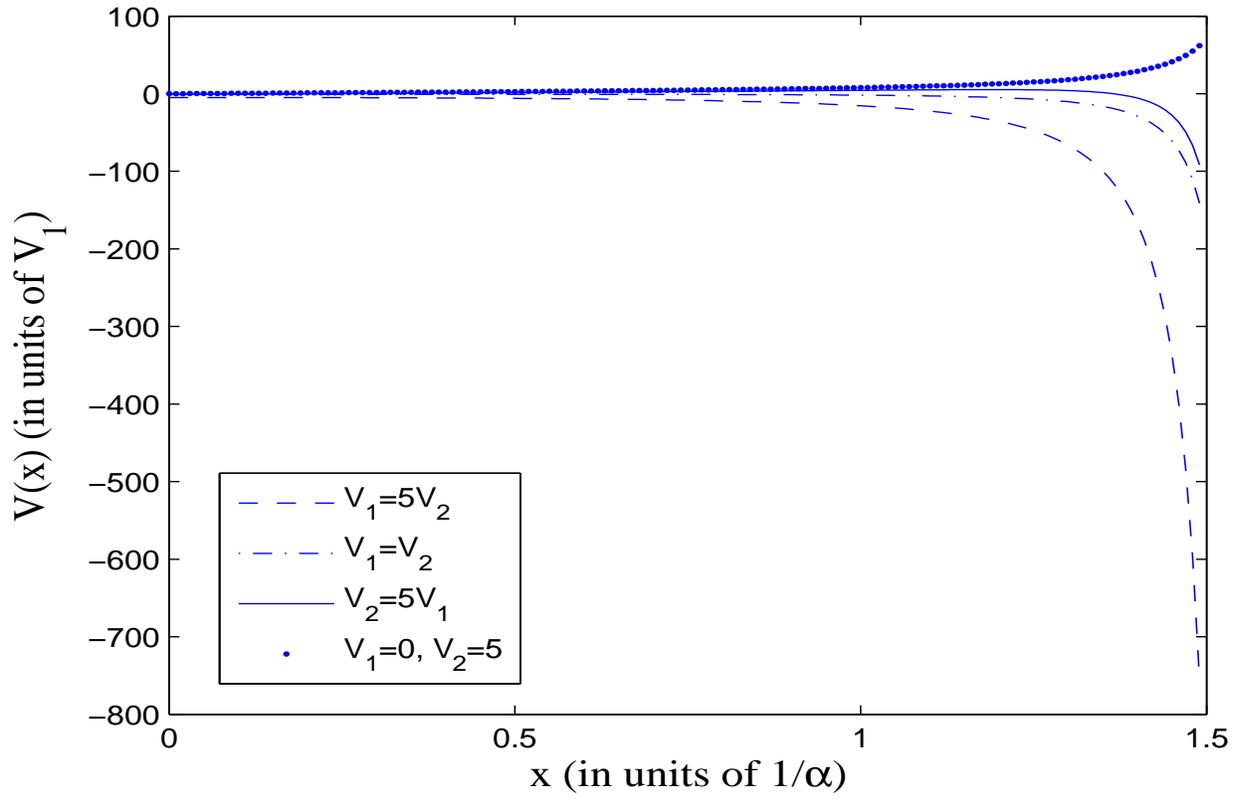} \caption{Plot of the trigonometric
Rosen-Morse-type potential [see Eq. (71)] for various sets of
parameters.}
\end{figure}

\end{document}